\documentclass[preprint,showpacs,preprintnumbers,amsmath,amssymb]{revtex4}

\usepackage{graphicx}

\begin{document}

\title{
Choreographic solution 
to the general relativistic three-body problem 
}
\author{Tatsunori Imai}
%\email{h05g302@stu.hirosaki-u.ac.jp}
\author{Takamasa Chiba}
%\email{h05gs310@stu.hirosaki-u.ac.jp}
\author{Hideki Asada} 
%\email{asada@phys.hirosaki-u.ac.jp}
\affiliation{
Faculty of Science and Technology, Hirosaki University,
Hirosaki 036-8561, Japan} 

\date{\today}

\begin{abstract}
We revisit the three-body problem in the framework of 
general relativity. 
The Newtonian N-body problem admits {\it choreographic} solutions, 
where a solution is called choreographic 
if every massive particles move periodically in a single closed orbit. 
One is a stable {\it figure-eight} orbit for a three-body system, 
which was found first by Moore (1993) and re-discovered 
with its existence proof by Chenciner and Montgomery (2000).  
In general relativity, however, the periastron shift prohibits 
a binary system from orbiting in a single closed curve. 
Therefore, it is unclear whether general relativistic effects  
admit a choreographic solution such as the figure eight. 
We carefully examine general relativistic corrections 
to initial conditions so that an orbit for a three-body system 
can be closed and a figure eight. 
This solution is still choreographic. 
This illustration suggests that the general relativistic N-body 
problem also may admit a certain class of choreographic solutions. 
\end{abstract}


\maketitle

%\section{Introduction}
\noindent \emph{Introduction.--- }
The three-body problem in the Newton gravity is one of classical problems 
in astronomy and physics (e.g, \cite{Danby}). 
In 1765, Euler found a collinear solution, 
and Lagrange found an equilateral triangle solution 
in 1772. 
It is impossible to describe all the solutions 
to the three-body problem even for the $1/r$ potential. 
In fact, Poincar\`e proved that we cannot analytically 
obtain all the solutions, 
and the number of new solutions is increasing \cite{Marchal}.   
Therefore, the three-body problem remains unsettled 
even for the Newtonian gravity. 
The Newtonian N-body problem admits {\it choreographic} solutions, 
which attract increasing interests. 
Here, a solution is called choreographic in the celestial mechanics 
if every massive particles move periodically in a single closed orbit. 
In fact, a choreographic figure-eight solution 
to the three-body problem
was found first by Moore \cite{Moore} and re-discovered 
with its existence proof by Chenciner and Montgomery \cite{CM}. 

The theory of general relativity is currently the most successful 
gravitational theory describing the nature of space and time, 
and well confirmed by observations. 
Especially, it has passed ``classical'' tests, 
such as the deflection of light, the perihelion shift 
of Mercury and the Shapiro time delay, and also a systematic test 
using the remarkable binary pulsar ``PSR 1913+16'' \cite{Will}. 
It is worthwhile to examine the three-body (or more generally, N-body) 
problem in general relativity. 
N-body dynamics in the general relativistic gravity 
plays important roles in astrophysics.  
For instance, the formation of massive black holes 
in star clusters is tackled mostly 
by Newtonian N-body simulations (e.g, \cite{MBH}). 
However, it is difficult to work out in general relativity 
compared with the Newtonian gravity, 
because the Einstein equation is much more complicated \cite{MTW} 
(even for a two-body system \cite{PW,FJS,Blanchet,AF}).  
In addition, future space astrometric missions 
such as SIM and GAIA 
\cite{SIM,GAIA,JASMINE}
require a general relativistic modeling of 
the solar system within the accuracy of a micro arc-second 
\cite{Klioner}. 
Furthermore, a binary plus the third body were discussed 
also for perturbations of gravitational waves induced by the third body 
\cite{ICTN,Wardell,CDHL,GMH}. 
In this paper, we do not intend to solve the N-body problem 
in general relativity under a general situation.  
Instead, we shall focus on a {\it choreographic} solution. 
No choreographic solution has not been found 
to the general relativistic N-body problem so far. 

In a two-body system, the post-Newtonian corrections cause 
the periastron shift so that the binary system cannot 
orbit in a single closed curve \cite{MTW}. 
As a result, it is unclear whether general relativistic perturbations 
admit a choreographic solution as the figure-eight. 
One may thus ask, ``what happens for the figure-eight 
in the Einstein's gravity?'' 
A specific question may arise, such as 
``does the figure-eight cause periastron shift?'',  
or 
``does the figure-eight make a transition to an open orbit 
in the general relativistic gravity?''  
The purpose of this paper is to answer these questions 
by carefully examining general relativistic effects to 
initial conditions for being a choreographic solution. 

This paper is organized as follows. 
First, we briefly summarize the choreographic 
figure-eight solution in the Newton gravity. 
Next, we analytically examine initial conditions and 
numerically solve the Einstein-Infeld-Hoffman 
equation of motion in order to obtain a choreographic solution 
in general relativity.  
Throughout this paper, we take the units of $G=c=1$. 

\begin{figure}
\includegraphics[width=10cm]{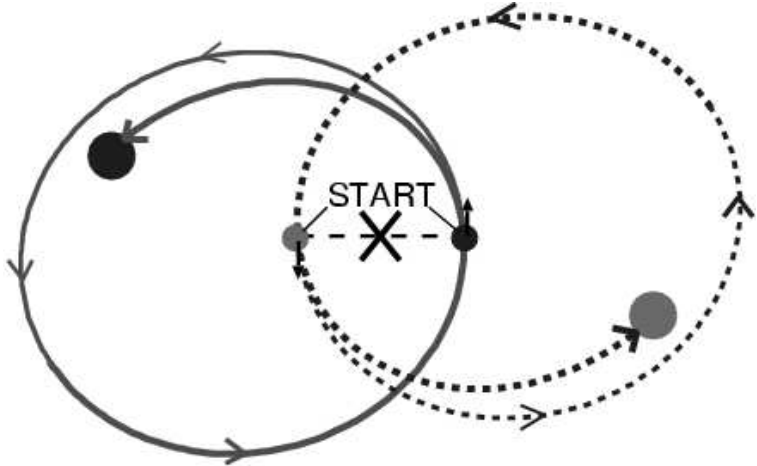}
\caption{
A schematic figure for a binary orbit in general relativity. 
The orbit is not closed any more, because of 
the periastron shift. 
}
\end{figure}

%\section{a choreographic solution in the Newton gravity} 
\noindent \emph{Newtonian choreographic solution.---} 
As mentioned 
above, 
it is impossible to describe all the solutions 
to the three-body problem even for the $1/r$ potential. 
The simplest periodic solutions for this problem 
were discovered by Euler (1765) and by Lagrange (1772). 
The Euler's solution is a collinear solution, 
in which the masses are collinear at every instant 
with the same ratios of their distances. 
The Lagrange's one is an equilateral triangle solution 
in which each mass moves in an ellipse in such a way 
that the triangle formed by the three bodies revolves.  
Built out of Keplerian ellipses, they are the only explicit
solutions. 
In these solutions, each mass moves on an ellipse. 
A choreographic solution for which three bodies move 
periodically in a single figure-eight orbit was 
found first by Moore by numerical computations \cite{Moore}. 
The existence of such a figure-eight orbit was proven 
by Chenciner and Montgomery \cite{CM}. 
This solution is stable in the Newtonian gravity 
\cite{Simo,GMF}. 
The figure-eight seems unique up to scaling 
and rotation according to all numerical investigations, 
and at the end its unicity has been recently proven 
\cite{Montgomery05}. 
Furthermore, it is shown numerically that fourth, 
sixth or eighth order polynomial cannot express 
the figure-eight solution \cite{Simo}. 
Nevertheless, no analytic expression in closed forms for 
the figure-eight trajectory has been found up to now.  
Therefore, in this paper, 
we numerically prepare the figure-eight orbit. 

\begin{figure}
\includegraphics[width=12cm]{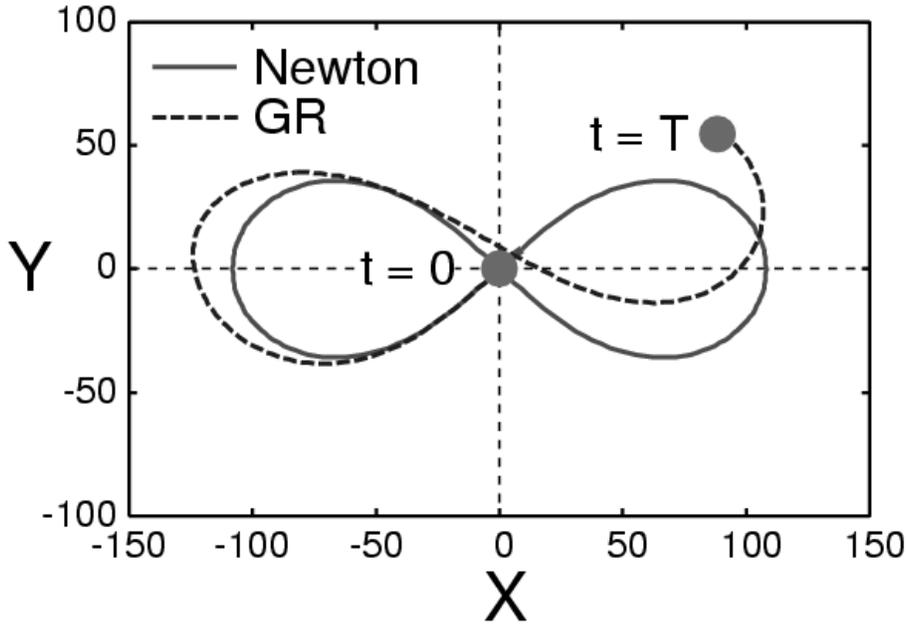}
\caption{
Figure-eights starting at the Newtonian initial condition. 
The solid curve denotes a figure-eight orbit 
in the Newtonian gravity. 
The dashed curve denotes a trajectory of one mass 
following the EIH equation of motion under  
the same Newtonian initial condition. 
}
\label{figure-eight1}
\end{figure}

For simplicity, we assume a three-body system with each mass 
equal to $m$. 
Without loss of the generality, the orbital plane 
is taken as the $x-y$ plane. 
The position of each mass ($m_A$) is denoted by 
$(x_A, y_A)$ for $A=1, 2, 3$. 
Figure $\ref{figure-eight1}$ shows the figure-eight orbit, 
where we take the initial condition 
as $\mbox{\boldmath $\ell$}\equiv (x_1, y_1)=(-x_2, -y_2)=
(97.00, -24.31)$, 
$(x_3, y_3)=(0, 0)$ 
and $\mbox{\boldmath $V$}_{\mbox{Newton}}\equiv  
(\dot{x}_3, \dot{y}_3)=(-2 \dot{x}_1, -2 \dot{y}_1)
=(-2 \dot{x}_2, -2 \dot{y}_2)
=(-0.09324, -0.08647)$ , 
where a dot denotes the time derivative \cite{Simo}. 
When one mass arrives at the knot (center) of the figure-eight, 
$\ell\equiv |\mbox{\boldmath $\ell$}|$ is 
a half of the separation between the remaining two masses. 
It is convenient to use $\ell$ instead of a distance between 
the knot and the apoapsis, because the inertial moment is 
expressed simply as $2 m \ell^2$. 
The orbital period is estimated as 
$
T_{\mbox{Newton}}=6.326 m^{-1/2} \ell^{3/2} 
\approx
10^4 
(M_{\odot}/m)^{1/2} 
(\ell/R_{\odot})^{3/2} 
\: 
\mbox{sec.} , 
$
where 
$M_{\odot}$ and $R_{\odot}$ are 
the solar mass and radius, respectively. 
Obviously this system has no Killing vector as seen 
in Fig. $\ref{figure-eight1}$. 
Here, we should note that $\ell$ is taken as 100, while 
it is the unity in the previous works. 
This is because we will treat the post-Newtonian correction 
in terms of the ratio between the mass and the separation 
such as $\ell$. 
In our case, the ratio $m/\ell$ is 0.01, that is, 
the post-Newtonian correction becomes about one percent. 
In the equation of motion, 
the second post-Newtonian (2PN) corrections of the order of 
$(m/\ell)^3$ can be safely neglected, if $m/\ell$ is very small, 
say, $10^{-8}$. 
In this case, however, the Newtonian and relativistic orbits 
will be indistinguishable. 
In order to demonstrate the difference between the two orbits, 
we choose $m/\ell$ as 0.01, for which the first post-Newtonian (1PN) 
terms are several dozens times larger than 2PN ones. 
In our computations, which are not long-time 
integrations over, say, thousands orbital periods, 
we can assume that 1PN terms are enough to bring 
major relativistic effects.

%\section{The post-Newtonian figure-eight solution}
\noindent \emph{Post-Newtonian figure-eight.---}
In the previous %section, 
part, 
the motion of massive bodies 
follows the Newtonian equation of motion. 
In order to include the dominant part of general relativistic effects, 
we take account of the terms at the first post-Newtonian order. 
Namely, the motion of the massive bodies obeys 
the Einstein-Infeld-Hoffman (EIH) equation of motion \cite{MTW}. 
The EIH equation is derived also from the first post-Newtonian 
Lagrangian as \cite{LL}
\begin{eqnarray}
{\cal L}&=& \frac12 \sum_A m_A v_A^2 
+ \frac12 \sum_A \sum_{B\neq A} \frac{m_A m_B}{r_{AB}} 
+ \frac18 \sum_A m_A v_A^4 
\nonumber\\
&&- \frac14 \sum_A \sum_{B\neq A} \frac{m_A m_B}{r_{AB}} 
\Biggl[
7 (\mbox{\boldmath $v$}_A \cdot \mbox{\boldmath $v$}_B)
-6 v_A^2 
+ (\mbox{\boldmath $v$}_A \cdot \mbox{\boldmath $n$}_{AB}) 
(\mbox{\boldmath $v$}_B \cdot \mbox{\boldmath $n$}_{AB}) 
\Biggr] 
\nonumber\\
&&- \frac12 \sum_A \sum_{B\neq A} \sum_{C\neq A} 
\frac{m_A m_B m_C}{r_{AB} r_{AC}} ,
\label{lagrangian}
\end{eqnarray}
where we define 
\begin{eqnarray}
\mbox{\boldmath $r$}_{AB} &\equiv& 
\mbox{\boldmath $r$}_{A}-\mbox{\boldmath $r$}_{B} , 
\\
r_{AB} &\equiv& |\mbox{\boldmath $r$}_{AB}| , 
\\
\mbox{\boldmath $n$}_{AB}&\equiv&
\frac{\mbox{\boldmath $r$}_{AB}}{r_{AB}} .
\end{eqnarray}

Figure $\ref{figure-eight1}$ shows an orbit of a body 
starting at the Newtonian initial condition described above. 
In Fig. $\ref{figure-eight1}$, a figure-eight orbit does not 
seem to survive at the 1PN order. 
However, this is not the case. 
We should note that 
the initial condition at the 1PN order does not necessarily  
coincide with that for the Newtonian gravity. 
We will thus carefully examine the initial condition 
by taking account of 1PN corrections. 
For this purpose, we assume that both the linear and angular 
momenta are zero ({\it i.e.}
$\mbox{\boldmath $P$} = 0$ and $\mbox{\boldmath $L$} = 0$).

We should remember 
$\mbox{\boldmath $v$}_1 = \mbox{\boldmath $v$}_2 = 
-(\mbox{\boldmath $v$}_3)/2$ 
for the Newtonian figure eight, 
for which both the total linear and angular momenta are zero. 
The changes in $\mbox{\boldmath $v$}_1$ and $\mbox{\boldmath $v$}_2$ 
are expressed by using two vectors $\mbox{\boldmath $v$}_3$ 
and $\mbox{\boldmath $\ell$}$, which are linearly independent.  

Hence, the initial velocity of each mass is parameterized as 
\begin{eqnarray}
\mbox{\boldmath $v$}_1&=&k\mbox{\boldmath $V$}+\xi 
\frac{m}{\ell ^3} 
(\mbox{\boldmath $V$}\cdot\mbox{\boldmath $\ell$})
\mbox{\boldmath $\ell$} , 
\label{v1}\\
\mbox{\boldmath $v$}_2&=&k\mbox{\boldmath $V$}+\xi
\frac{m}{\ell ^3} 
(\mbox{\boldmath $V$}\cdot\mbox{\boldmath $\ell$}) 
\mbox{\boldmath $\ell$} , 
\label{v2}\\ 
\mbox{\boldmath $v$}_3&=&\mbox{\boldmath $V$} , 
\label{v3}
\end{eqnarray}
where $k$ is expressed as 
\begin{equation}
k=-\frac{1}{2}+\alpha |\mbox{\boldmath $V$}|^2+\beta \frac{m}{\ell} . 
\end{equation}
Here, the 1PN terms have either $|\mbox{\boldmath $V$}|^2$ 
or $m/\ell$. 
If $\mbox{\boldmath $P$} = 0$ and $\mbox{\boldmath $L$} = 0$, 
there is no need of $|\mbox{\boldmath $V$}|^2$ in front of 
$\mbox{\boldmath $\ell$}$ in Eqs. ($\ref{v1}$) and ($\ref{v2}$) 
as shown below. 

The linear and angular momenta are calculated from 
the first post-Newtonian Lagrangian \cite{LL}. 
Here, we impose the condition of $\mbox{\boldmath $P$} = 0$ and 
$\mbox{\boldmath $L$} = 0$ at 1PN order.  
Then, we determine the 1PN coefficients as 
\begin{eqnarray}
\alpha&=&-\frac{3}{16} ,
\\
\beta&=&\frac{1}{8} , 
\\
\xi&=&\frac{1}{8} . 
\end{eqnarray}
Ut to this point, $\mbox{\boldmath $V$}$ is arbitrary. 
Next, we determine $\mbox{\boldmath $V$}$. 

The initial velocity of the particles can be different from 
that for the Newtonian gravity. 
The post-Newtonian effects affect both the magnitude and direction 
of the velocity. 
Therefore, by using two linearly independent vectors, 
$\mbox{\boldmath $\ell$}$ and $\mbox{\boldmath $V$}_{\rm Newton}$, 
we parameterize the initial velocity as 
\begin{equation}
\mbox{\boldmath $V$}=\left(1+\delta 
\frac{m}{\ell}\right)\mbox{\boldmath $V$}_{\rm Newton}
+\eta \frac{m}{\ell}\frac{\mbox{\boldmath $\ell$}}{\ell}
\left(\mbox{\boldmath $V$}_{\rm Newton}\cdot
               \frac{\mbox{\boldmath $\ell$}}{\ell}\right) , 
\label{V}
\end{equation}
where it is sufficient to express 1PN corrections in terms of 
either $m/\ell$ or $|\mbox{\boldmath $V$}_{\rm Newton}|^2$ 
in numerical computations, though both are necessary 
for analytic calculations of 
$\mbox{\boldmath $P$} = 0$ and $\mbox{\boldmath $L$} = 0$. 
For convenience's sake, we choose $m/\ell$ in Eq. ($\ref{V}$). 

By numerically performing trial and error iterations until achieving 
a periodic orbit, we find out 
\begin{eqnarray}
\delta &=&-3.3 ,
\label{delta}\\
\eta &=& -3.7 .    
\label{eta}
\end{eqnarray}
Here, the iterative computations are done until 
we find the values of $\delta$ and $\eta$ for which 
the three masses simultaneously return to their initial positions. 
Our procedure is as follows. 
If and only if one particle returns to the neighborhood of 
its initial position ({\it i.e.} the origin for the particle 
labeled by 3) within the positional deviation of $0.01$, 
we measure how far the remaining two particles are from 
their initial positions at the same moment when 
the particle is closest to the initial position. 
For Eqs. $(\ref{delta})$ and $(\ref{eta})$, 
the sum of the square distances is minimized as approximately 0.1. 
This is sufficient for $\ell=100$ and $m/\ell=0.01$, 
because expected positional shifts after one cycle 
are of the order of the unity or more. 
For instance, such a shift exceeds ten for 
$\alpha = \beta =\xi = \delta = \eta = 0$ 
in Fig. $\ref{figure-eight1}$. 

We integrate the motion over ten cycles 
to confirm the periodicity. 
After ten periods, the found solution comes to 
the same point within the deviation of $\pm 1$. 
The numerical computation gives the orbital period as 
\begin{equation}
T_ {\mbox{GR}}\approx\left(1+\frac{6m}{\ell}\right) 
\times T_{\mbox{Newton}} . 
\end{equation}
The relativistic figure eight appears to be stable, 
in the sense that we recognize 'eight-like' orbits that 
resemble figure eight 
for slightly different values of $\delta$ and $\eta$. 
It is a future subject to analyze the long-time stability 
of the relativistic figure eight. 

\begin{figure}
\includegraphics[width=12cm]{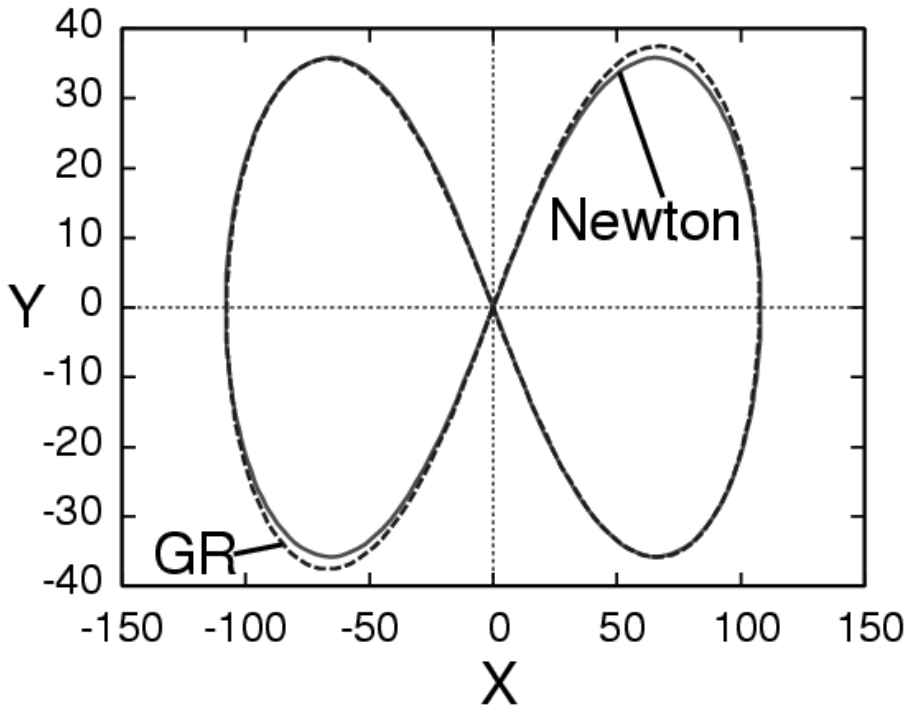}
\caption{
Figure-eight orbits. 
The solid curve denotes a figure-eight orbit in the Newtonian
gravity. The dashed curve denotes a figure-eight orbit 
at the 1PN order of general relativity. 
}
\label{fig-PN8}
\end{figure}

\begin{figure}
\includegraphics[width=12cm]{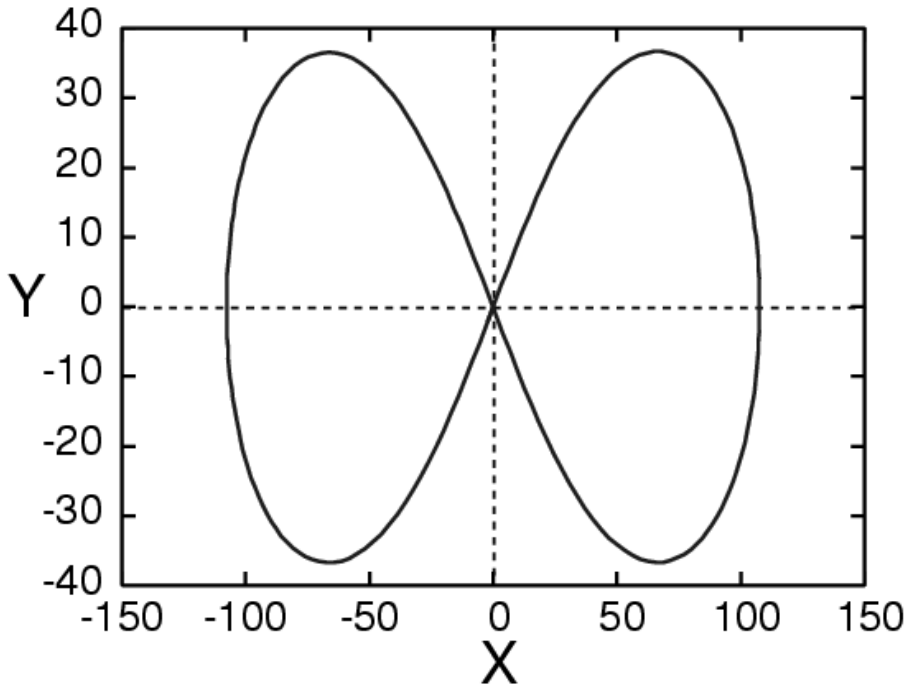}
\caption{
Relativistic figure eight. 
The principal axes of the orbit are chosen as $x$ and $y$ axes.   
}
\label{fig-PN8-2}
\end{figure}

Figure $\ref{fig-PN8}$ shows that a figure-eight orbit 
is still closed even after including the dominant general 
relativistic effects. 
In Fig. $\ref{fig-PN8}$, we can recognize 
an asymmetric difference between the Newtonian figure-eight orbit and 
the general relativistic (GR) one. 
The deviation is partly fiducial, because 
the principal axes of the GR figure-eight orbit are 
not along the $x$ and $y$ axes. 
That is, $\mbox{\boldmath $\ell$}$, which defines 
the direction of the initial position of a particle 
with respect to the principal axes, 
changes slightly at the 1PN order. 
The axes are inclined by 0.012 radian 
with respect to those of the Newtonian figure-eight orbit. 
Now, we choose the $x$-axis as the principal axis for both 
the Newtonian figure-eight orbit and the general relativistic one. 
After choosing the principal axes, 
Fig. $\ref{fig-PN8-2}$ shows a general relativistic choreographic 
solution at the first post-Newtonian order. 
The solution recovers line symmetry with respect to 
the $x$ and $y$ axes. 
There are no significant differences in the velocity 
between the Newtonian and GR figure-eight orbits. 
One may notice that the 2PN terms are neglected. 
It would be safer to choose $m/\ell$ for instance as $10^{-8}$ 
and then on the figures to exaggerate the differences 
between the Newtonian and post-Newtonian solutions 
in order to make them visible.

Finally, we mention the possibility 
of three-body systems in a choreographic orbit such as a figure-eight. 
As a new outcome of binary-binary scattering, the figure-eight
orbit was discussed for presenting a way of detecting such an orbit 
in numerical computations \cite{Heggie}. 
According to the numerical result, the probability of the formation 
of figure-eight orbits is a tiny fraction of one percent. 
The gravitational waves emitted by the figure-eight 
have been recently studied by assuming the motion in 
the Newton gravity \cite{GW8}. 
By evaluating the radiation reaction time scale, 
it is shown also that figure-eight sources emitting 
gravitational waves may be too rare to detect.  

%\section{Conclusion}
\noindent \emph{Conclusion.--- }
We obtained a general relativistic initial condition 
for being a figure-eight orbit.  
This condition provides the first choreographic solution 
taking account of the post-Newtonian corrections. 
It is interesting to include higher post-Newtonian 
corrections, especially 2.5PN effects in order to 
elucidate the backreaction on the evolution of the orbit 
due to the gravitational waves emission at the 2.5PN order. 
If the system is secularly stable against the gravitational radiation, 
one might see probably a shrinking ($\dot\ell < 0$) figure-eight orbit 
as a consequence of a decrease in the total energy ($\dot E<0$). 
This speculation will be confirmed or rejected in future. 
It may be important also to look for other relativistic 
choreographic solutions for a system including four or more masses. 
It is possible that some of Newtonian choreographic solutions are 
prohibited by general relativistic effects. 
Further investigations along these lines will allow us 
to probe many-body dynamics in the Einstein gravity.


\begin{thebibliography}{99}
\bibitem{Danby}
J. M. A. Danby, {\it Fundamentals of Celestial Mechanics} 
(William-Bell, VA, 1988) 
\bibitem{Marchal}
C. Marchal, {\it The Three-Body Problem} 
(Elsevier, Amsterdam, 1990) 
\bibitem{Moore} 
C. Moore, 
Phys. Rev. Lett. {\bf 70}, 3675 (1993). 
\bibitem{CM} 
A. Chenciner, R. Montgomery, 
Ann. Math. {\bf 152}, 881 (2000). 
\bibitem{Will}
C. M. Will, {\it Theory and experiment in gravitational physics} 
(Cambridge Univ. Press, Cambridge, 1993) 
\bibitem{MBH}S. P. Zwart, H. Baumgardt, P. Hut, 
J. Makino, S. McMillan, Nature {\bf 428}, 724 (2004).
\bibitem{MTW}
C. W. Misner, K. S. Thorne, J. A. Wheeler, 
{\it Gravitation}, 
(Freeman, New York, 1973).
\bibitem{PW}
M. E. Pati, C. M. Will, 
Phys. Rev. D {\bf 65} 104008  (2002). 
\bibitem{FJS}
G. Faye, P. Jaranowski, G. Schafer, 
Phys. Rev. D {\bf 69} 124029 (2004). 
\bibitem{Blanchet}
L. Blanchet, 
Living Rev.  {\bf 9}, 4 (2006). 
\bibitem{AF}
H. Asada, T. Futamase, 
Prog. Theor. Phys. Suppl. No. {\bf 128}, 123 (1997). 
\bibitem{SIM}
http://planetquest.jpl.nasa.gov/SIM/sim$\underline{\;}$index.cfm.
\bibitem{GAIA}
http://www.rssd.esa.int/index.php?project=GAIA\&page=index.
\bibitem{JASMINE}
http://www.jasmine-galaxy.org/index.html.
\bibitem{Klioner}
S. A. Klioner, 
Astron. J.  {\bf 125} 1580 (2003). 
\bibitem{ICTN}
K. Ioka, T. Chiba, T. Tanaka, T. Nakamura, 
Phys. Rev. D {\bf 58}, 063003 (1998). 
\bibitem{Wardell}
Z. E. Wardell, 
Mon. Not. R. Astron. Soc. {\bf 334}, 149 (2002). 
\bibitem{CDHL}
M. Campanelli, M. Dettwyler, M. Hannam, C. O. Lousto, 
Phys. Rev. D {\bf 74}, 087503 (2006). 
\bibitem{GMH}
K. Gultekin, M. C. Miller, D. P. Hamilton, 
Astrophys.J. {\bf 640} 156 (2006).
\bibitem{GMF}J. Galan, F. J. Munoz-Almaraz, E. Freire, 
E. Doedel, A. Vanderbauwhede, 
Phys. Rev. Lett. {\bf 88}, 241101 (2002). 
\bibitem{Simo}
C. Simo, 
Contemp. Math. {\bf 292}, 209 (2002). 
\bibitem{Montgomery05} 
R. Montgomery, 
Ergodic Theory and Dynamical Systems {\bf 25}, 921 (2005). 
\bibitem{LL}L. D. Landau and E. M. Lifshitz, {\it The Classical Theory 
of Fields} (Oxford: Pergamon 1962). 
\bibitem{Heggie}
D. C. Heggie, 
Mon. Not. R. Astron. Soc. {\bf 318}, L61 (2000). 
\bibitem{GW8}
T. Chiba, T. Imai, H. Asada, 
accepted for publication in 
Mon. Not. R. Astron. Soc., (astro-ph/06097773).
\end{thebibliography}
\end{document}